\documentclass{ws-procs10x7}
\usepackage{balance}
\usepackage{psfig}
\usepackage{pennames}

\newcolumntype{d}[1]{D{.}{.}{#1}}

\def\Journal#1#2#3#4{{\it #1} {\bf #2}, #3 (#4)}

\makeindex
\begin{document}

\title{WW PRODUCTION, POLARISATION \\ AND SPIN CORRELATIONS AT LEP}

\author{P. AZZURRI}

\address{Scuola Normale Superiore, Piazza dei Cavalieri 7, 56126 Pisa, Italy
\\E-mail: paolo.azzurri@sns.it}

\twocolumn[\maketitle

\abstract{
Results on W-pair productions at LEP2 are reviewed and their consistency with
standard electroweak expectations is summarised. 
Special interest is given to measurements of W polarisation,
and to new studies of spin and decay-plane correlations among the W pairs.
}
]

\section{Introduction}
During the data taking years from 1996 to 2000 the LEP2 program 
has delivered $\Pep\Pem$ interactions at centre of mass energies 
ranging from 161 to 207~GeV.
In this period the four LEP experiments (ALEPH, DELPHI, L3, OPAL)
have collected a total integrated luminosity of almost 3~fb$^{-1}$.  
The LEP2 data yielded about 40,000 W pair events, reconstructed in all
their decay channels. 
Event rates allowed measurements of the W pair partial and total cross sections
at different energies, as well as of W decay branching fractions. 
Further, the study of the 
W pair decay products kinematics allowed determinations of  
the W polarisation states, and of correlations among the two 
W bosons.

\section{W pair production rates}
\begin{figure}[hbt]
\centerline{\psfig{file=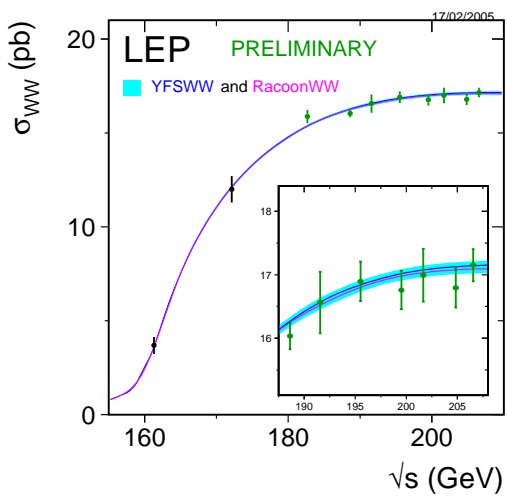,width=7cm}}
\caption{Total W pair cross section in $\Pep\Pem$ collisions measured at LEP2 
at centre of mass energies from 161 to 207 GeV. Dots with error bars are the 
experimental measurements while the continuos band represents
the theoretical predictions based on the Standard Model. \label{figxs}
}
\end{figure}

Depending on the decay channels, different techniques allowed  to select W pairs 
with efficiencies in the 50-90\% range and background 
contamination in the 5-20\% range~\cite{ew05,Axs,Dxs,Lxs}.
The combined results for the total W pair cross section at LEP2 energies 
are shown in figure~\ref{figxs} where the combined experimental precision
is at the level of 1\%. The total cross section results represent the first clear 
proof of the electroweak SU(2)$\otimes$U(1) gauge structure, through the 
gauge cancellations that ensure the W pair
tree level scattering unitarity and the theory renomalizability.

The cross section measurement near the W pair production kinematic threshold 
at $\sqrt{s}$=161~GeV, has allowed a determination of the 
W boson mass, $m_\PW=80.40\pm0.21$~GeV/c$^2$, 
 that is completely independent from the methods based on the kinematics of the 
decay products.
The overall experimental precision of the cross section measurements has also allowed 
to establish the presence of Standard Model radiative corrections to the tree level
CC03 diagrams involved in the W pair production. 
Depending on the energy, these radiative 
corrections decrease the total cross section at the 2-3\% level, and modify
the shapes of some differential cross section distributions.

The W pair data has also allowed measurements of the W decay branching fractions
at the 1\% level. Combined results~\cite{ew05} are in good agreement with the Standard Model
expectations, except for the tau leptonic branching fraction that appears to exceed 
both the electron and muon fractions by more than two standard deviations leading to
$2g_\tau/(g_{\rm e}+g_\mu)= 1.036\pm 0.014$, 
an overall excess of 2.6 standard deviations. 

\section{W Polarisation}
The helicity of the produced W can be measured using the polar and 
azimuthal decay angles ($\theta^\ast, \phi^\ast$) of the decay fermion 
in the W rest frame.
In the case of a leptonic decay, where the charge of the fermion can be 
determined, the W boson fractions of helicity states are given by
\begin{eqnarray*}
\frac1N\frac{dN}{d\cos\theta^\ast_\ell}=
f_-\frac38 \left( 1+ \cos\theta^\ast_\ell\right)^2 +\\
f_+\frac38 \left( 1- \cos\theta^\ast_\ell\right)^2 +
f_0\frac34\sin^2\theta^\ast_\ell ,
\end{eqnarray*}
where $f_-$, $f_0$, $f_+$ are the fractions of  the 
$\lambda=-1,0,+1$ helicity states for a \PWm, and of the 
$\lambda=+1,0,-1$ helicity states for a \PWp. 
In the case of hadronic decays, since the charge can't be determined clearly,
only the absolute value $|\cos\theta^\ast_\Pq|$ is measurable and the differential 
distribution becomes
\begin{eqnarray*}
\frac1N\frac{dN}{d|\cos\theta^\ast_\Pq|}=
f_\pm\frac34 \left( 1+ |\cos\theta^\ast_\Pq |^2\right) +\\
f_0\frac34 \left( 1- |\cos\theta^\ast_\Pq |^2 \right) ,
\end{eqnarray*}
where $f_\pm=f_++f_-$ is the sum of both transverse helicity fractions.
 
In the case of LEP pair productions, the helicity state will be a function of the 
centre of mass energy $s$ and of the angle $\theta_\PWm$ of the \PWm 
direction with respect to the $\Pep\Pem$ axis. 
To measure the helicity fractions, the interference terms  and possible CP or CPT
violating effects, the formalism of the Spin Density Matrix (SDM) is used,
defined as
$$
\sigma^{\PWm}_{\tau\tau'}(s,\cos\theta_{\PWm})=
F_\tau^{\PWm} (F_{\tau'}^{\PWm})^\ast / \sum_\tau |F_\tau^{\PWm}|^2
$$ 
where $F_\tau^{\PWm}$ is the amplitude to produce a W$^-$
with helicity $\tau=-1,0,+1$.
The SDM is a Hermitian matrix with unit trace, the diagonal elements
are the probabilities to observe a W in one of the three helicity states 
 and the off diagonal terms represent the interference between helicity 
 amplitudes. 

\begin{figure}[hbt]
\centerline{\psfig{file=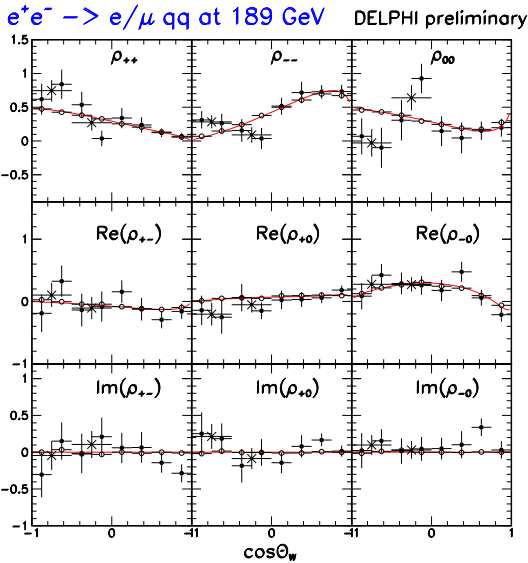,width=7cm}}
\caption{Measurements of the Spin Density Matrix elements
from DELPHI~\protect\cite{Dp}. 
The values are shown as a function of $\cos\theta_\PWm$ 
averaging $\PWm$ and $\PWp$ data at $\sqrt{s}$=
189~GeV.
Full lines and open circles represent Standard Model prediction calculations,
while the black dots and crosses are the data values.\label{figsdm}}
\end{figure}

LEP experiments have measured SDM elements at different energies and
 in different $\cos\theta_{\PW}$ bins~\cite{Lp,Op,Dp}, by measuring the average
values of projection operators $\Lambda^{\PWm}_{\tau\tau'}$ that are
analytical functions of  the $\theta^\ast$ and  $\phi^\ast$ angles.
 An example of SDM measurements is shown in figure~\ref{figsdm}.

Invariance under CPT  would ensure that 
   $\sigma^{\PWm}_{\tau\tau'}=\left( \sigma^{\PWp}_{-\tau-\tau'}\right)^\ast$
while additional CP invariance implies 
 $ \sigma^{\PWm}_{\tau\tau'}=\sigma^{\PWp}_{-\tau-\tau'}$
so that the matrix would be real.
Therefore effects that do not conserve CP or CPT have been searched for by comparing 
the  $\PWp$ and $\PWm$ SMD elements, and by searching for 
off diagonal imaginary parts.
No significant deviation from the Standard Model expectation has been seen in the data
so that both the CP and CPT violating effects in the SDM should
be at a level smaller that 10$^{-1}$.   

Focusing on the diagonal SDM elements, the overall fraction of longitudinally
polarised W bosons at LEP2 can be extracted averaging $\sigma_{00}$ over all data samples.
The average of current results~\cite{Lp,Op,Dp} yields
$$
f_L=\frac{\sigma_{00}}{\sigma_{\rm tot}}=23.6\pm1.6\%.
$$

\section{Spin and Decay-Plane Correlations}
Recent studies address the possibility to measure 
correlations between the spin states of the two W bosons 
in W pair events, and more in general correlations in the decay kinematics 
of the two W bosons~\cite{Lc}.

\begin{figure}[hbt]
\centerline{\psfig{file=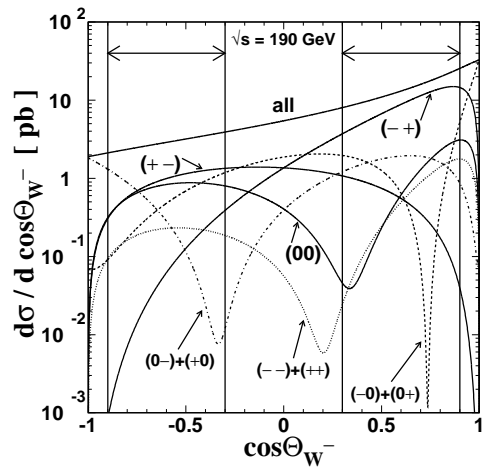,width=6.5cm}
}
\caption{Differential $\cos\theta_{\PWm}$ cross section distribution
for production of polarised W pairs at $\sqrt{s}$=190~GeV, from ref.~\protect\cite{Lc}.
The figure shows the contribution to the total cross section by each possible pair 
of $\PWm$ and $\PWp$ helicities, indicated in parenthesis.
The two  $\cos\theta_{\PWm}$  intervals indicated by the arrows show the
backward region (low $\cos\theta_{\PWm}$) with increased contributions from (0,0) longitudinal
 helicities  and the forward region (high $\cos\theta_{\PWm}$) with increased contributions from (-,+)
 transverse helicities.\label{figc1}}
\end{figure}

The electroweak model predicts that the helicity combinations of the two W bosons
depend strongly on the direction of the production axis $\cos\theta_\PWm$ as shown
in figure~\ref{figc1}.
The forward $\cos\theta_{\PWm}$ direction is where the maximum of W pairs, 
both with transverse helicity ($\lambda_\PWm=-1,\lambda_\PWp=+1$) is 
found, and selecting $0.3<\cos\theta_\PWm<0.9$, 68\% of the W pairs are expected to be
produced in the (-,+) state.    
Conversely in the backward $\cos\theta_{\PWm}$ direction the maximum of W pairs, 
with double longitudinal helicity ($\lambda_\PWm=0,\lambda_\PWp=0$) is 
expected, in particular a fraction of 28\% in the $-0.9<\cos\theta_\PWm<-0.3$ region.

Spin correlations have been searched for in semileptonic 
$\PWp\PWm\rightarrow\Pq\Paq\ell\nu$ decays. 
For the W decaying hadronically a cut 
$|\cos\theta^\ast_\Pq|>0.66$ on the polar decay angle 
distribution allows to select a  
sample enriched in $\lambda=\pm 1$ transverse spin while a cut
$|\cos\theta^\ast_\Pq|<0.33$ selects a sample of hadronic decays 
enriched in longitudinal $\lambda=0$ spin.
For both samples a fit to the lepton $\cos\theta^\ast_\ell$ 
distribution determines the helicity fractions 
of the leptonic W, and differences in the fitted fractions 
can reveal the correlations among the two spin states.

\begin{figure}[hbt]
\centerline{\psfig{file=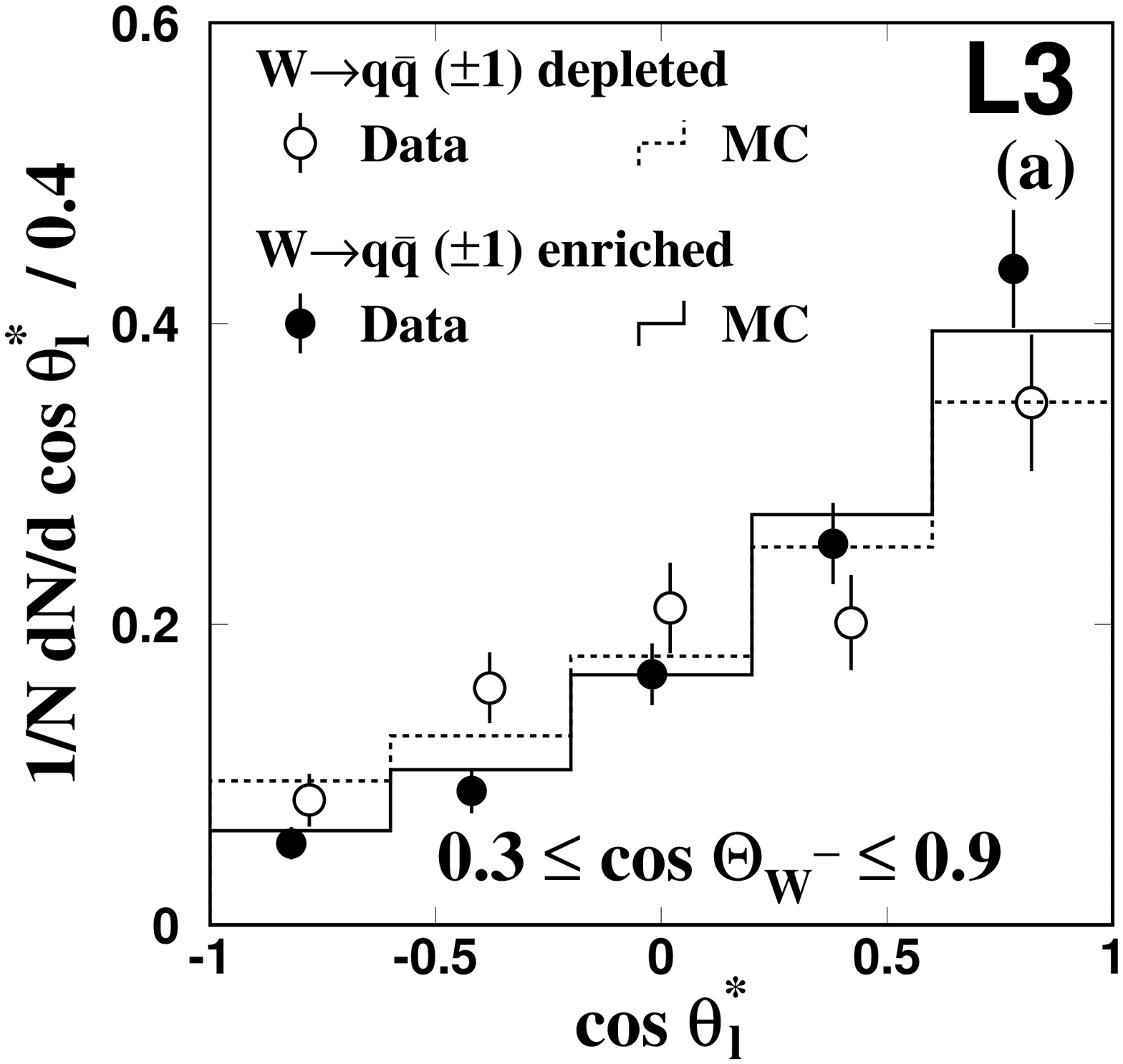,width=3.8cm}
\psfig{file=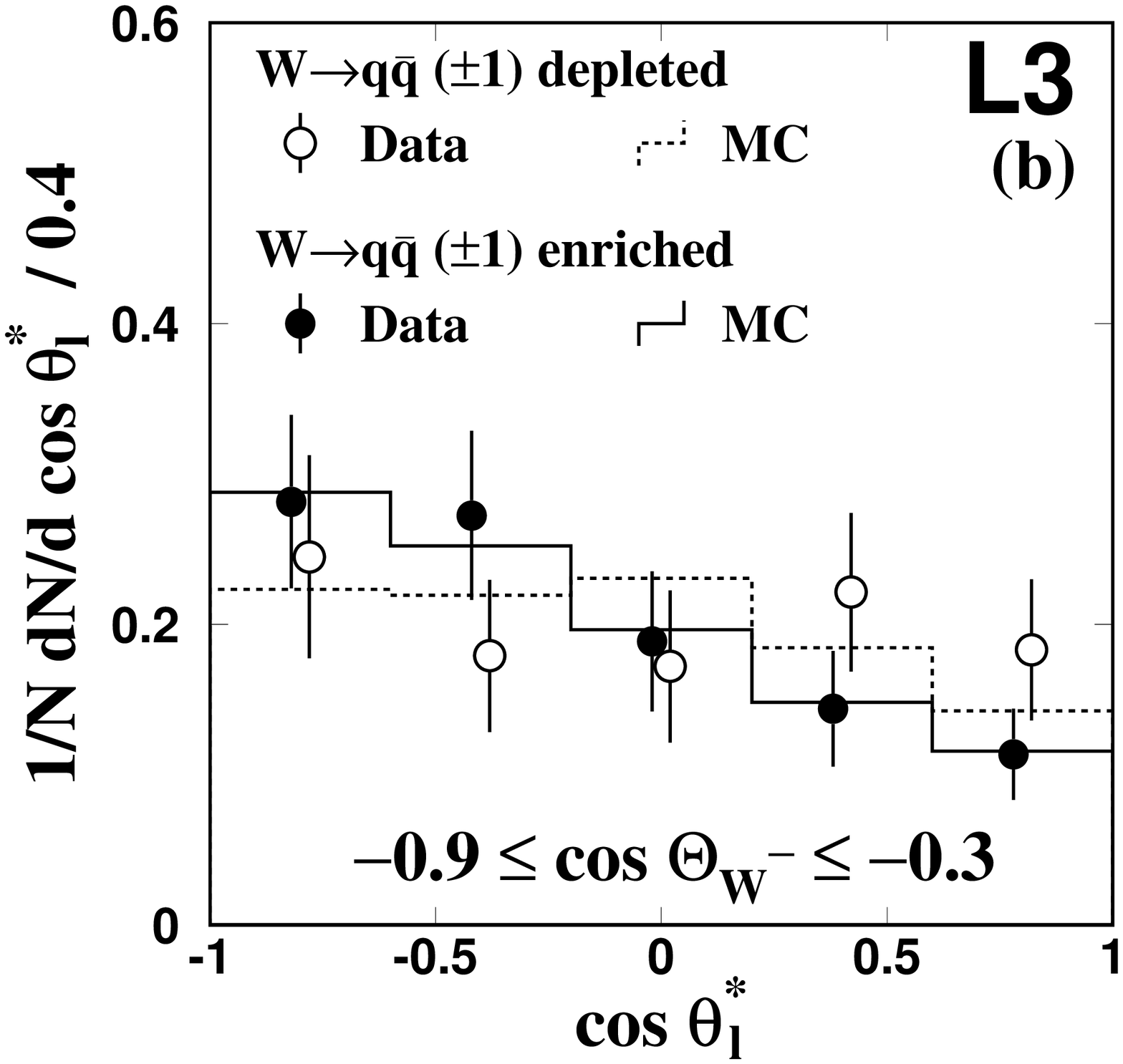,width=3.8cm}
}
\caption{Distribution of $\cos\theta^\ast_\ell$ 
from L3~\protect\cite{Lc} $\PW\rightarrow\ell\nu$ decays, 
given in the intervals
$0.3<\cos\theta_\PWm<0.9$ (a) and 
$-0.9<\cos\theta_\PWm<-0.3$ (b). Data points and Monte~Carlo
predictions are shown for two samples, enriched and depleted 
in transverse helicity for the other W boson in the event. 
\label{figc4}
}
\end{figure}

In the forward $\cos\theta_\PWm$ region, enriching the hadronic transverse 
spin states should enhance the leptonic transverse spin fractions, while 
in the backward $\cos\theta_\PWm$ region the enrichment of hadronic
longitudinal states should enhance the leptonic longitudinal spin fraction.
Data results are shown in figure~\ref{figc4} where the expected effects are 
visible in the forward $\cos\theta_\PWm$ data, which benefits from quite 
larger statistics. 

With the forward $\cos\theta_\PWm$ data, enriching 
the hadronic W transverse spin, with
the two $|\cos\theta^\ast_\Pq|$ selections, changes the leptonic W helicity fractions
by $\Delta f_-=+0.32\pm0.10\pm0.06$, 
$\Delta f_+=-0.03\pm0.07\pm0.05$ and  $\Delta f_0=-0.28\pm0.14\pm0.08$,
clearly  enhancing the transverse helicities, in reasonable agreement with the 
Standard Model expectations.
\begin{figure}[hbt]
\centerline{\psfig{file=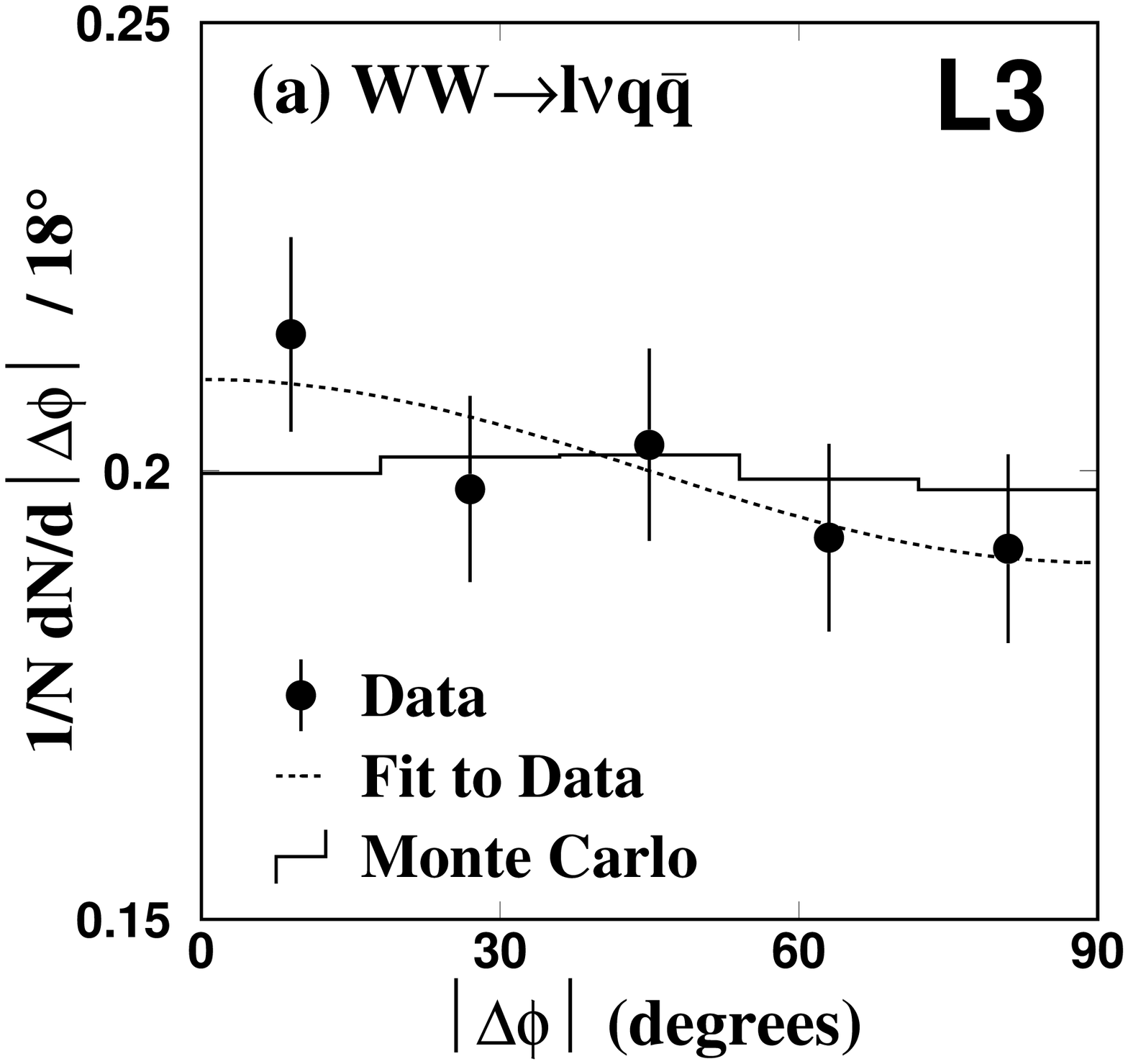,width=3.8cm}
\psfig{file=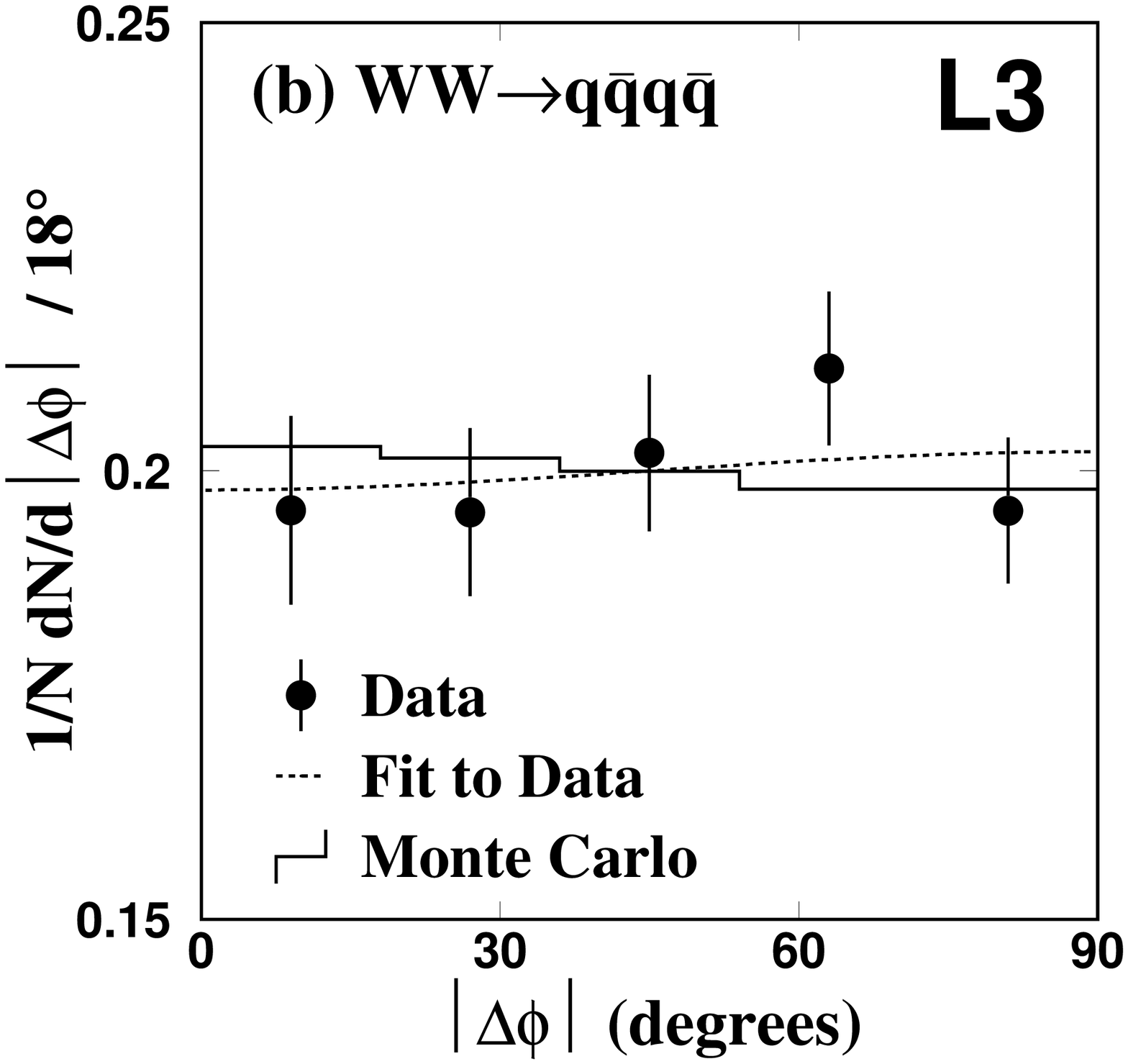,width=3.8cm}
}
\caption{Distribution of the angle $|\Delta\phi|$ between
the decay planes of the two W bosons from (a)
$\PWp\PWm\rightarrow\ell\nu\Pq\Paq$ events
and (b) $\PWp\PWm\rightarrow\Pq\Paq\Pq\Paq$ events from L3~\protect\cite{Lc}.
The Monte~Carlo predictions are shown with the data points 
and the results of the fit to the data.
\label{figc5}
}
\end{figure}

In a more general approach both semileptonic and fully hadronic W pair decays have been
used to determine possible correlations between the the two W bosons decay planes.
Given the angle $|\Delta\phi|$ between the planes of the decay products of the two W bosons, 
the differential distribution of $|\Delta\phi|$ is given by
$$
\frac1N\frac{dN}{d|\Delta\phi|}= 1+ D\cos |\Delta\phi|
$$
where $D$ measures the strength of the correlation.
The data distributions of $|\Delta\phi|$ are shown in figure~\ref{figc5}. 
The combined result of the fits of the two distributions gives 
$D=0.012\pm0.021\pm0.012$, in agreement with the Standard Model 
Monte~Carlo prediction of $D_{MC}=0.010\pm0.002$, but not revealing any 
significant correlation effect.

\section{Conclusions}
The LEP2 data has brought the first precision confirmation of the
SU(2)$\otimes$U(1) electroweak gauge structure through the W pair productions
rates and decay kinematics. Results are in good agreement with the 
Standard Model expectations, with the $\PW\tau\Pgngt$ coupling 2.6 standard
deviations larger than expected.
The W polarization fractions have been measured accurately and no anomalous interference terms, CP or CPT violating contributions have been found in the Spin Density matrix determinations. 
Spin correlations have also been established between the two W bosons,
in agreement with  expectations.

\end{document}